\documentstyle[twoside,epsfig]{article}
\catcode`\@=11
\long\def\@makefntext#1{
\protect\noindent \hbox to 3.2pt {\hskip-.9pt  
$^{{\eightrm\@thefnmark}}$\hfil}#1\hfill}		%CAN BE USED 

\def\@makefnmark{\hbox to 0pt{$^{\@thefnmark}$\hss}}	%ORIGINAL 
	
\def\ps@myheadings{\let\@mkboth\@gobbletwo
\def\@oddhead{\hbox{}
\rightmark\hfil\eightrm\thepage}   
\def\@oddfoot{}\def\@evenhead{\eightrm\thepage\hfil
\leftmark\hbox{}}\def\@evenfoot{}
\def\sectionmark##1{}\def\subsectionmark##1{}}

\oddsidemargin=\evensidemargin
\newcounter{sectionc}\newcounter{subsectionc}\newcounter{subsubsectionc}
\renewcommand{\section}[1] {\vspace{12pt}\addtocounter{sectionc}{1} 
\setcounter{subsectionc}{0}\setcounter{subsubsectionc}{0}\noindent 
	{\tenbf\thesectionc. #1}\par\vspace{5pt}}
\renewcommand{\subsection}[1] {\vspace{12pt}\addtocounter{subsectionc}{1} 
	\setcounter{subsubsectionc}{0}\noindent 
	{\bf\thesectionc.\thesubsectionc. {\kern1pt \bfit #1}}\par\vspace{5pt}}
\renewcommand{\subsubsection}[1] {\vspace{12pt}\addtocounter{subsubsectionc}{1}
	\noindent{\tenrm\thesectionc.\thesubsectionc.\thesubsubsectionc.
	{\kern1pt \tenit #1}}\par\vspace{5pt}}

\newcounter{appendixc}
\newcounter{subappendixc}[appendixc]
\newcounter{subsubappendixc}[subappendixc]
\renewcommand{\thesubappendixc}{\Alph{appendixc}.\arabic{subappendixc}}
\renewcommand{\thesubsubappendixc}
	{\Alph{appendixc}.\arabic{subappendixc}.\arabic{subsubappendixc}}

\renewcommand{\appendix}[1] {\vspace{12pt}
        \refstepcounter{appendixc}
        \setcounter{figure}{0}
        \setcounter{table}{0}
        \setcounter{lemma}{0}
        \setcounter{theorem}{0}
        \setcounter{corollary}{0}
        \setcounter{definition}{0}
        \setcounter{equation}{0}
        \renewcommand{\thefigure}{\Alph{appendixc}.\arabic{figure}}
        \renewcommand{\thetable}{\Alph{appendixc}.\arabic{table}}
        \renewcommand{\theappendixc}{\Alph{appendixc}}
        \renewcommand{\thelemma}{\Alph{appendixc}.\arabic{lemma}}
        \renewcommand{\thetheorem}{\Alph{appendixc}.\arabic{theorem}}
        \renewcommand{\thedefinition}{\Alph{appendixc}.\arabic{definition}}
        \renewcommand{\thecorollary}{\Alph{appendixc}.\arabic{corollary}}
        \renewcommand{\theequation}{\Alph{appendixc}.\arabic{equation}}
        \noindent{\tenbf Appendix \theappendixc #1}\par\vspace{5pt}}
\newcommand{\subappendix}[1] {\vspace{12pt}
        \refstepcounter{subappendixc}
        \noindent{\bf Appendix \thesubappendixc. {\kern1pt \bfit #1}}
	\par\vspace{5pt}}
\newcommand{\subsubappendix}[1] {\vspace{12pt}
        \refstepcounter{subsubappendixc}
        \noindent{\rm Appendix \thesubsubappendixc. {\kern1pt \tenit #1}}
	\par\vspace{5pt}}

\newcommand{\textlineskip}{\baselineskip=13pt}
\newcommand{\smalllineskip}{\baselineskip=10pt}

\def\eightcirc{
\begin{picture}(0,0)
\put(4.4,1.8){\circle{6.5}}
\end{picture}}
\def\eightcopyright{\eightcirc\kern2.7pt\hbox{\eightrm c}} 

\newcommand{\copyrightheading}[1]
	{\vspace*{-2.5cm}\smalllineskip{\flushleft
	{\footnotesize International Journal of Modern Physics A, #1}\\
	{\footnotesize $\eightcopyright$\, World Scientific Publishing
	 Company}\\
	 }}

\def\abstracts#1#2#3{{
	\centering{\begin{minipage}{4.5in}\baselineskip=10pt\footnotesize
	\parindent=0pt #1\par 
	\parindent=15pt #2\par
	\parindent=15pt #3
	\end{minipage}}\par}} 

\renewenvironment{thebibliography}[1]
	{\frenchspacing
	 \ninerm\baselineskip=11pt
	 \begin{list}{\arabic{enumi}.}
	{\usecounter{enumi}\setlength{\parsep}{0pt}
	 \setlength{\leftmargin 12.7pt}{\rightmargin 0pt} %FOR 1--9 ITEMS
	 \setlength{\itemsep}{0pt} \settowidth
	{\labelwidth}{#1.}\sloppy}}{\end{list}}

\newcounter{itemlistc}
\newcounter{romanlistc}
\newcounter{alphlistc}
\newcounter{arabiclistc}

\newcommand{\fcaption}[1]{
        \refstepcounter{figure}
        \setbox\@tempboxa = \hbox{\footnotesize Fig.~\thefigure. #1}
        \ifdim \wd\@tempboxa > 5in
           {\begin{center}
        \parbox{5in}{\footnotesize\smalllineskip Fig.~\thefigure. #1}
            \end{center}}
        \else
             {\begin{center}
             {\footnotesize Fig.~\thefigure. #1}
              \end{center}}
        \fi}

\newcommand{\tcaption}[1]{
        \refstepcounter{table}
        \setbox\@tempboxa = \hbox{\footnotesize Table~\thetable. #1}
        \ifdim \wd\@tempboxa > 5in
           {\begin{center}
        \parbox{5in}{\footnotesize\smalllineskip Table~\thetable. #1}
            \end{center}}
        \else
             {\begin{center}
             {\footnotesize Table~\thetable. #1}
              \end{center}}
        \fi}

\def\@citex[#1]#2{\if@filesw\immediate\write\@auxout
	{\string\citation{#2}}\fi
\def\@citea{}\@cite{\@for\@citeb:=#2\do
	{\@citea\def\@citea{,}\@ifundefined
	{b@\@citeb}{{\bf ?}\@warning
	{Citation `\@citeb' on page \thepage \space undefined}}
	{\csname b@\@citeb\endcsname}}}{#1}}

\newif\if@cghi
\def\cite{\@cghitrue\@ifnextchar [{\@tempswatrue
	\@citex}{\@tempswafalse\@citex[]}}
\def\citelow{\@cghifalse\@ifnextchar [{\@tempswatrue
	\@citex}{\@tempswafalse\@citex[]}}
\def\@cite#1#2{{$\null^{#1}$\if@tempswa\typeout
	{IJCGA warning: optional citation argument 
	ignored: `#2'} \fi}}

\def\pmb#1{\setbox0=\hbox{#1}
	\kern-.025em\copy0\kern-\wd0
	\kern.05em\copy0\kern-\wd0
	\kern-.025em\raise.0433em\box0}

\def\fnt#1#2{\footnotetext{\kern-.3em
	{$^{\mbox{\scriptsize #1}}$}{#2}}}

\def\fpage#1{\begingroup
\voffset=.3in
\thispagestyle{empty}\begin{table}[b]\centerline{\footnotesize #1}
	\end{table}\endgroup}

\def\runninghead#1#2{\pagestyle{myheadings}
\markboth{{\protect\footnotesize\it{\quad #1}}\hfill}
{\hfill{\protect\footnotesize\it{#2\quad}}}}
\font\tenrm=cmr10
\font\tenit=cmti10 
\font\tenbf=cmbx10
\font\bfit=cmbxti10 at 10pt
\font\ninerm=cmr9

\font\eightrm=cmr8

\def\qed{\hbox{${\vcenter{\vbox{			%HOLLOW SQUARE
   \hrule height 0.4pt\hbox{\vrule width 0.4pt height 6pt
   \kern5pt\vrule width 0.4pt}\hrule height 0.4pt}}}$}}

\begin{document}

\runninghead{Instructions for Typesetting Camera-Ready
Manuscripts $\ldots$} {Instructions for Typesetting Camera-Ready
Manuscripts $\ldots$}

\normalsize\textlineskip
\thispagestyle{empty}
\setcounter{page}{1}

\copyrightheading{}			%{Vol. 0, No. 0 (1993) 000--000}

\vspace*{0.88truein}

\fpage{1}
\centerline{\bf REAL TIME NONEQUILIBRIUM DYNAMICS OF QUANTUM PLASMAS.}
\vspace*{0.035truein}
\centerline{\bf  QUANTUM KINETICS AND THE DYNAMICAL RENORMALIZATION
GROUP.}
\vspace*{0.37truein}
\centerline{\footnotesize H. J. de VEGA}
\vspace*{0.015truein}
\centerline{\footnotesize\it LPTHE, Universit\'e 
Paris VI et Paris VII, Tour 16, 1er. \'etage, 4,
Place Jussieu} 
\baselineskip=10pt
\centerline{\footnotesize\it 75252 Paris cedex 05, FRANCE}
\vspace*{10pt}
\vspace*{0.21truein}
\abstracts{We implement the  dynamical renormalization group (DRG) using the 
hard thermal loop (HTL) approximation for the real-time nonequilibrium
dynamics in hot plasmas. The focus is on the study of the
relaxation of gauge and fermionic mean fields and on the quantum
kinetics of the photon and fermion distribution functions.
As a concrete physical prediction,
we find that for a QGP of temperature $T \sim 200\;\mbox{MeV}$
and lifetime $10\leq t \leq 50\;\mbox{fm}/c$
there is a new contribution to the hard ($k \sim T$) photon production 
from off-shell bremsstrahlung
($q\rightarrow q\gamma$ and $\bar{q}\rightarrow\bar{q}\gamma$)
at just ${\cal O}(\alpha)$ that grows logarithmically in time and is
{\bf comparable} to the known on-shell Compton scattering
and pair annihilation at ${\cal O}(\alpha\,\alpha_s)$. }{}{}

%\textlineskip			%) USE THIS MEASUREMENT WHEN THERE IS
%\vspace*{12pt}			%) NO SECTION HEADING

\vspace*{1pt}\textlineskip	%) USE THIS MEASUREMENT WHEN THERE IS
%\section{General Appearance}	%) A SECTION HEADING
%\vspace*{-0.5pt}
\vspace*{0.5truecm}
\noindent
The study of nonequilibrium phenomena under extreme conditions
play a fundamental r\^ole in the understanding of ultrarelativistic heavy
ion collisions and early universe cosmology. 

Thermal field theory provides the tools to study the properties of
plasmas in {\em equilibrium}~\cite{revius},
but the consistent study of nonequilibrium phenomena in real time requires
the methods of nonequilibrium field theory~\cite{noneq} (see
also Refs.~\cite{chalo} for further references). The study of the
equilibrium and nonequilibrium properties of abelian and non-abelian
plasmas as applied to the QGP has as ultimate goal a deeper understanding
of the potential experimental signatures of the formation and evolution of
the QGP in ultrarelativistic heavy ion collisions.
Amongst these, photons and dileptons (electron and/or muon pairs)
produced during the early stages of the QGP  are considered as some of
the most promising signals~\cite{photdilep,kapusta}. Since
photons and lepton pairs interact electromagnetically their mean free paths are
longer than the estimated size of the QGP fireball $\sim 10-50\;
\mbox{fm}$ and unlike hadronic signals they do not undergo final state
interactions. Therefore photons and  dileptons produced during the
early stages of QGP carry clean information from this phase.

{ \bf The goals our programme}.
aim to provide a comprehensive study of several
relevant aspects of the {\em nonequilibrium} dynamics of abelian QED
plasma as well as nonabelian QCD plasma in {\bf real time}. 

This programme about the {\bf real time} evolution of field theory
started in 1995 studying scalar theories \cite{nos1} and was later generalized
to hot scalar QED \cite{nos2}. Our starting point is the
Schwinger-Dyson equation for the expectation value of the field. For
example, for the spatial Fourier transform of the transverse gauge
field in QED $ {\bf a}_T({\vec x},t) $,
\begin{equation}
\left(\frac{\partial^2}{\partial t^2}+k^2 \right) \; {\bf a}_T({\vec k},t) +
\int^{t}_{-\infty}dt'\;\Pi_T({\vec k},t-t') \;
{\bf a}_T({\vec k},t')=0 \; ,\label{maxwell}
\end{equation}
and we neglect non-linear contributions of the order $ {\cal O}\left(
{ a}_T\right)^2 $. Here $\Pi_T({\vec k},t-t')$ is the transverse
part of the retarded photon self-energy.

Eq.(\ref{maxwell}) can be readily solve by Laplace transform with
solution\cite{nos1,nos2,nos3}
\begin{equation}
\tilde{\bf a}_T(s,{\vec k})=\frac{1}{s}\left[1-
{ k^2 \over s^2+k^2+ \tilde{\Pi}_T(s,{\vec k})}\right]\;
{\bf a}_T({\vec k},0) \label{laplasol}\;,
\end{equation}
where
$$
\tilde{\bf a}_T(s,{\vec k})=\int_0^{\infty} dt\; e^{-st}\;  {\bf
a}_T({\vec k},t)\; .
$$
We focus on the following\cite{nos3}:
\begin{itemize}
\item[(i)]{The real time evolution of gauge mean fields in linear
response in the HTL approximation. The goal here is to study
directly in real time the relaxation of (coherent) gauge field configurations
in the linearized approximation to leading order in the HTL
program. The analogous  study for  {\em scalar} QED is given in
ref.\cite{nos2}} 

\item[(ii)]{The quantum kinetic equation that describes the evolution of the
distribution function of photons in the medium, again to leading order
in the HTL approximation. This aspect is relevant to study photon
production via off-shell effects directly in real time. 
This quantum kinetic equation, obtained from a microscopic
field theoretical approach based on the dynamical renormalization
group~\cite{drg} (DRG) displays novel off-shell effects that
cannot be captured via the usual kinetic description that assumes
completed collisions~\cite{photdilep,kapusta}.}

\item[(iii)]{The evolution in real time of fermionic mean fields
features anomalous relaxation arising from the emission and
absorption of magnetic photons (gluons) which are only dynamically
screened by Landau damping~\cite{robinfra,blaizotBN}.
The Bloch-Nordsieck approximation for the fermion propagator
provides a resummation of the
infrared divergences associated with soft photon (or gluon)
bremsstrahlung in the medium~\cite{blaizotBN}. In ref.\cite{nos3}
 we implement the DRG to study the
evolution of fermionic mean fields providing an  alternative to the
Bloch-Nordsieck treatment.}

\item[(iv)]{We obtain in ref.\cite{nos3} the quantum kinetic equation
for the fermionic distribution function for hard fermions via the 
DRG. The DRG leads to a quantum kinetic equation
directly in real time bypassing the assumption of completed collisions
and leads to a {\em time-dependent} collision
kernel free of infrared divergences.}
\end{itemize}

%\subsection{\bf Relaxation of gauge mean fields}
\noindent
We studied the
relaxation of a gauge mean field in linear response to leading
order in the HTL approximation both for soft momentum $ k \leq  e T$ and
for semihard momentum  $ eT \ll k \ll T $ under the assumption of
weak electromagnetic coupling.

{\em Soft momentum} ($k \sim eT$): in this case
 the relaxation of the gauge mean field is
dominated by the end-point contribution of the Landau damping cut.
As a consequence, the soft gauge mean field relaxes with a power law
long time tail of the form
$$
{\bf a}_T({\vec k},t) \buildrel{kt\gg 1}\over=
{\bf a}_T({\vec k},0)
\left[\frac{k^2 Z_T(k)}{\omega^2_T(k)}\cos[\omega_T(k)t]
-\frac{12}{e^2 T^2}\frac{\cos kt}{t^2}\right]\; ,
$$
where $\omega_T(k)$ is the transverse photon pole and $Z_T(k)$
is the corresponding residue.
We note that in spite of the power law tail the gauge mean
field relaxes towards the oscillatory mode determined by the
transverse photon pole. This reveals that the soft collective excitation
in a plasma is stable in the HTL approximation.

{\em Ultrasoft momentum} ($ k \ll  eT$): In the region of ultrasoft momentum
the spectral density divided by the frequency  features a sharp
Breit-Wigner peak near zero 
frequency in the region of Landau damping, with width $\Gamma_k =
12k^3/\pi e^2 T^2 $. We find that the amplitude of a mean field of
transverse photons is given by 
$$
{\bf a}_T({\vec k},t)\buildrel{kt\gg 1\, , \, \Gamma_k \, t \leq
1}\over= {\bf a}_T({\vec k},0) 
\left[\frac{k^2 Z_T(k)}{\omega^2_T(k)}\cos[\omega_T(k)t]+
e^{-\Gamma_k t}\right]\; .
$$

{\em Semi-hard momentum} ($ eT \ll k \ll T $):
In this region both the HTL approximation and the perturbative
expansion are formally valid. However the spectral density in the
Landau damping region is sharply peaked near 
$\omega = k$ and the transverse photon pole approaches the
edge of the Landau damping region from above. Although
the perturbative expansion is in principle valid, the
sharp spectral density near the edge of the continuum results in a
breakdown of the perturbative expansion. The DRG
provides a consistent resummation  of the lowest order HTL
perturbative contributions in real time, leading for the relaxation at
intermediate asymptotic times:
$$
{\bf a}_T({\vec k},t) \buildrel{kt \gg 1}\over= {\cal A}_T({\vec
k},t) \left( \frac{t}{\tau_0} \right)^{-\frac{e^2T^2}{12k^2}}\;,
$$
where $\tau_0\sim 1/k$ and ${\cal A}_T({\vec k},t)$ is an
oscillating function. 
The anomalous exponent is a consequence of an infrared enhancement
arising from the sharp spectral density near the threshold of the
Landau damping region for semihard momentum. The crossover to
exponential relaxation due to collisional processes at higher orders
is discussed in \cite{nos3}.

\noindent
Using the techniques of nonequilibrium
field theory and the DRG, we obtain the
quantum kinetic equation for the distribution
function of semihard photons $eT\ll k \ll T$ to lowest order in the HTL
approximation assuming that the fermions are thermalized. An
important result is that the collision kernel is {\em time-dependent}
and the DRG reveals that detailed balance
emerges during microscopic time scales, i.e,
much shorter than the relaxation scales. In the linearized
approximation we find that the departure from equilibrium
of the photon distribution function relaxes as:
$$
\delta n^\gamma_{\vec k}(t)=\delta n^\gamma_{\vec k}(t_0)
\left(\frac{t-t_0}{\tau_0}\right)^{-\frac{e^2 T^2}{6k^2}}
\quad\mbox{for}\quad k(t-t_0)\gg 1\; ,
$$
where $\tau_0\sim 1/k$, and $t_0$ is the initial time.
Furthermore, this quantum kinetic equation
allows us to study photon production by off-shell
effects, which to leading order in the HTL approximation 
is of order $\alpha$.
Extrapolating the result from QED to thermalized QGP with two
flavors of light quarks, we find that the total number of hard
photons at time $t$ per invariant phase space volume to lowest order is
\begin{equation}\label{alpha}
N(t) =  \frac{5  \alpha T^3}{18 \pi^2 k^2} \left\{\ln \left[
2k(t-t_0)\right]+\gamma_E-1\right\} \quad \mbox{for} \quad k(t-t_0) > 1\;.
\end{equation}
with $t_0 \approx 1\mbox{fm/c}$ is the time scale at which the QGP
plasma is thermalized. 

We find that for a quark-gluon plasma at temperature $T \sim 200\;\mbox{MeV}$
and of lifetime $10 \leq (t-t_0) \leq 50\;\mbox{fm}/c$, this new
hard ($k\sim T$) photon production by off-shell bremsstrahlung is comparable to
the known contribution from Compton scattering and pair annihilation of 
order $\alpha\,\alpha_s$~\cite{kapusta}.
This is a noteworthy result and our  {\bf main point}: we find
photon production to lowest (one-loop) order arising solely from
{\bf off-shell} effects:
($q\rightarrow q\gamma$ and $\bar{q}\rightarrow\bar{q}\gamma$) and
annihilation of quarks ($q\bar{q} \rightarrow\gamma$).

\noindent
We implement the DRG resummation to study the
real-time relaxation of a fermion mean field for hard momentum. The
emission and absorption of magnetic photons which
are only dynamically screened by Landau damping introduce a logarithmic
divergence in the spectral density near the fermion mass shell.
The DRG resums these
divergences in real time and leads to a relaxation of the
fermion mean field for hard momentum given by:
$$
\psi({\vec k},t)\buildrel{kt \gg 1}\over= e^{-\alpha
Tt\left[\ln(\omega_P t)+0.12652\ldots \right]} \times
\mbox{oscillating phases},
$$
with $\omega_P$ being the plasma frequency.

We obtain a quantum kinetic equation for the
distribution function of hard fermions using non-equilibrium field
theory and the DRG resummation. In the linearized approximation the
distribution function relaxes as:
$$
\delta n^f_{\vec k}(t)\buildrel{kt \gg 1}\over=
\delta n^f_{\vec k}(t_0)\; e^{-2\alpha T(t-t_0)\,
\left[\ln(\omega_P t)+0.12652\ldots\right]}\;,
$$
where the anomalous relaxation exponent is twice that of the mean field.

\begin{figure}[htbp]
\epsfig{file=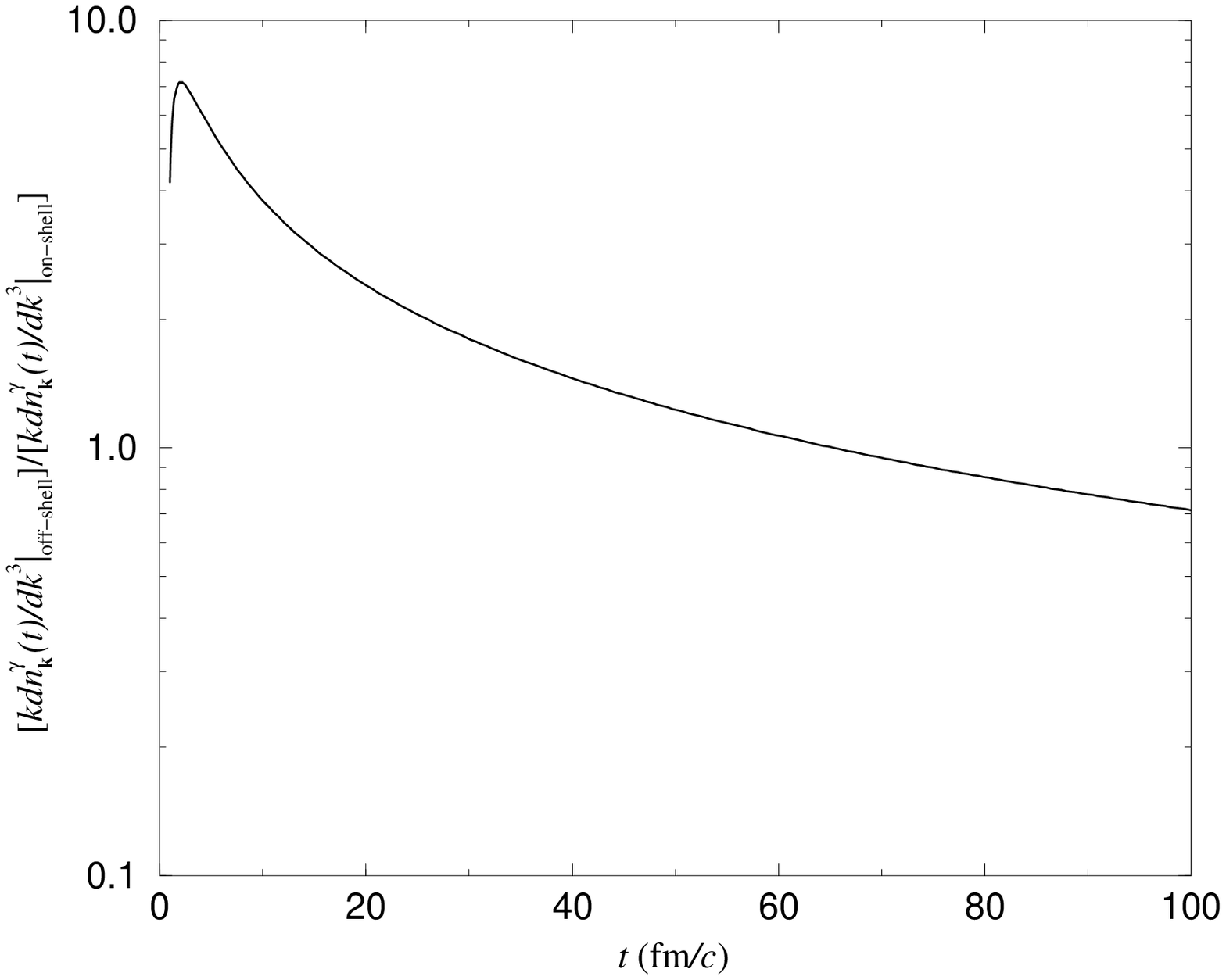,width=12.0cm,height=8.0cm}
\fcaption{The ratio of the number of hard photons
produced by offshell processes according to eq.(\ref{alpha})
($q\rightarrow q\gamma$ and 
$\bar{q}\rightarrow\bar{q}\gamma$) to that produced by on-shell
processes ($qg\rightarrow q\gamma$ and $q\bar{q}\rightarrow
g\gamma$) as a function of QGP lifetime $t$ for $\alpha=1/137$,
$\alpha_s=0.4$ and $T=200\;{\rm MeV}$. \label{fig:photonratio}}
\end{figure}


\begin{thebibliography}{000}

\bibitem{revius}
M. Le Bellac, {\em Thermal Field Theory}, Cambridge University Press, 1996.
H. A. Weldon, Phys. Rev. {D28}, 2007 (1983); Ann. of Phys. (N.Y.)
{\bf 228}, 43 (1993); 

\bibitem{noneq}
J. Schwinger, J. Math. Phys. {\bf 2}, 407 (1961),
K. T. Mahanthappa, Phys. Rev. {\bf 126}, 329 (1962);
P. M. Bakshi and K.T. Mahanthappa, J. Math. Phys. {\bf 41}, 12 (1963),
L. V. Keldysh, JETP {\bf 20}, 1018 (1965).

\bibitem{chalo} D. Boyanovsky, H.J. de Vega and R. Holman, in Proceedings of
the Second Paris Cosmology Colloquium, Observatoire de Paris, 1994,
edited by H.J. de Vega and N. S\'anchez,
World Scientific, Singapore 1995, pp. 127-215;
in {\em Advances in Astrofundamental Physics},
Erice Chalonge School, edited by N. S\'anchez and
A. Zichichi, World Scientific, Singapore 1995.

\bibitem{photdilep}
L. McLerran and T. Toimela,  Phys. Rev. {\bf D31},
545, (1985); P. V. Ruuskanen, Nucl. Phys. {\bf A544}, 169c (1995).
\bibitem{kapusta}
J. Kapusta, P. Lichard and D. Seibert,
Phys. Rev.  {\bf D44}, 2774 (1991).
R. Baier, M. Dirks, K. Redlich and D. Schiff,
Phys. Rev. {\bf D56}, 2548 (1997);
R. Baier, H. Nakkagawa, A. Niegawa and K. Redlich,
Z.  Phys.  C {\bf 53}, 433 (1992);
R. Baier, B. Pire and D. Schiff,
Phys. Rev.{\bf D38}, 2814 (1988).
\bibitem{nos1} D. Boyanovsky, M. D'Attanasio,  H. J. de Vega,
R. Holman and  D.-S. Lee, Phys. Rev. {\bf D52}, 6805 (1995).  
D. Boyanovsky, H. J. de Vega, M. D'Attanasio and  R. Holman,
Phys. Rev. {\bf D54}, 1748 (1996).
D. Boyanovsky,  H. J. de Vega, Y. J. Ng, D.-S. Lee and
S.-Y. Wang, Phys. Rev. {\bf D59}, 105001 (1999)
J. Baacke, D. Boyanovsky,  H. J. de Vega, hep-ph/9907337, to appear in
Phys. Rev. {\bf D}. 

\bibitem{nos2} D. Boyanovsky, H. J. de Vega, R. Holman, S. Prem Kumar
and  R. D. Pisarski, Phys. Rev. {\bf D58}, 125009 (1998). 
D. Boyanovsky, H. J. de Vega, D.-S. Lee and  Y. J. Ng and
S.-Y. Wang, Phys. Rev. {\bf D61}, 065004 (2000).

\bibitem{nos3}
D. Boyanovsky, H. J. de Vega, D.-S. Lee and S.-Y. Wang,
Phys. Rev. {\bf D62}, 105026 (2000). 

\bibitem{drg}D. Boyanovsky, H. J. de Vega, R. Holman and
M. Simionato,  Phys. Rev. {\bf D60}, 065003 (1999).
D. Boyanovsky, H. J. de Vega, Y. J. Ng, D.-S. Lee and
S.-Y. Wang, Phys. Rev. {\bf D59}, 105001 (1999).
D. Boyanovsky, H. J. de Vega, Phys. Rev. {\bf D59}, 105019 (1999).
D. Boyanovsky, H. J. de Vega and S.-Y. Wang,
Phys. Rev. {\bf D61}, 065006 (2000).

\bibitem{robinfra}
R.D. Pisarski, Phys. Rev. Lett. {\bf 63}, 1129 (1989);
Phys. Rev. D {\bf 47},5589 (1993).
\bibitem{blaizotBN}
J.-P. Blaizot and E. Iancu, Phys. Rev. Lett. {\bf 76}, 3080 (1996);
Phys. Rev. D {\bf 55}, 973 (1997); Phys. Rev. D {\bf 56}, 7877 (1997).
K. Takashiba, Int. J. Mod. Phys {\bf A 11}, 2309 (1996).
\end{thebibliography}
\end{document}